\documentclass[a4paper,12pt]{article}
\usepackage[DIV13]{typearea}
\usepackage{amsmath} 
\usepackage{amssymb}
\usepackage{amsfonts}
\usepackage{amscd}
\usepackage{graphicx}
\usepackage{epsfig}
\usepackage[footnotesize,bf]{caption2}
\usepackage{cite}
\usepackage{bbold}
\usepackage{longtable}
\usepackage[center,footnotesize,hang]{subfigure}
\usepackage{url}

\title{Non-Abelian Discrete Flavor Symmetries from $T^2/Z_N$ Orbifolds}
\author{A. Adulpravitchai \footnote{E-mail: \texttt{adisorn.adulpravitchai@mpi-hd.mpg.de}}~, A. Blum \footnote{E-mail: \texttt{alexander.blum@mpi-hd.mpg.de}}~, and M. Lindner \footnote{E-mail: \texttt{manfred.lindner@mpi-hd.mpg.de}} \\\\
{\normalsize \it Max--Planck--Institut f\"ur Kernphysik,}\\
{\normalsize \it  Postfach 103980, D--69029 Heidelberg, Germany}}
\date{}

\begin{document}

\maketitle

\begin{abstract}
In \cite{GA} it was shown how the flavor symmetry $A_4$ (or $S_4$) can arise if the three fermion generations are taken to live on the fixed points of a specific 2-dimensional orbifold. The flavor symmetry is a remnant of the 6-dimensional Poincar\'e symmetry, after it is broken down to the 4-dimensional Poincar\'e symmetry through compactification via orbifolding. This raises the question if there are further non-abelian discrete symmetries that can arise in a similar setup. To this end, we generalize the discussion by considering all possible 2-dimensional orbifolds and the flavor symmetries that arise from them. The symmetries we obtain from these orbifolds are, in addition to  $S_4$ and $A_4$, the groups $D_3,D_4$ and $D_6 \simeq D_3 \times Z_2$ which are all popular groups for flavored model building.
\end{abstract}

\section{Introduction}

The flavor problem of the Standard Model of Particle Physics has two aspects. First, the question arises what flavor is. Next, one can ask why the parameters of the flavor sector, the fermion masses and the mixing matrices, take the values they do. A popular and successful approach is to impose a non-abelian discrete flavor symmetry to explain certain observed regularities. The nature of flavor is, in the context of flavor symmetries, therefore usually reduced to the question as to the origin of that symmetry.

Two main types of symmetries are needed to construct the Lagrangian of the Standard Model: space-time and gauge symmetries. In general adding an additional gauge group to the Standard Model is a much simpler task than extending the space-time symmetry. However, breaking a continuous flavor gauge group down to a non-abelian discrete subgroup is a highly non-trivial phenomenological task. In particular, for such a breaking, large representations of the continuous symmetry are needed, which can not couple directly to the small representations in which the three generations of fermions would reside\footnote{For further details we refer to \cite{nogo}.}. It is thus worthwhile to consider discrete flavor symmetries arising as extensions of the space-time symmetry.

An extension of the space-time symmetry can only be achieved by an extension of space-time itself. We thus need to work in an extra-dimensional framework. Such an extension of space-time will enlarge the Poincar\'e symmetry. If the $n$ extra dimensions are compactified in an orbifold, the space-time symmetry will not be the full $4+n$-dimensional Poincar\'e symmetry. However, depending on the exact compactification, there may be residual discrete symmetries, which can then play the role of flavor symmetry.

This idea was first explored in \cite{GA}, where two extra dimensions were assumed. This can be considered the minimal number in this setup, as one extra dimension does not lead to non-abelian symmetries. For a specific 2-dimensional orbifold it was shown there, that the residual Poincar\'e symmetry is the group $S_4$, the group of permutations of four distinct objects (if discrete symmetries, such as parity, are not taken into account, i.e. if we only consider proper Lorentz transformations, the residual symmetry is $A_4$). $A_4$ \cite{A4papers} and $S_4$ \cite{S4papers} are both popular and phenomenologically successful as flavor symmetries, especially for predicting tri-bimaximal neutrino mixing. In this paper we generalize the discussion of \cite{GA} by considering all possible 2-dimensional orbifolds and calculating the resulting symmetry. As it turns out, the resulting flavor symmetries are, in addition to $A_4$ and $S_4$, the three dihedral groups $D_4$, $D_3 \cong S_3$ and $D_6 \cong D_3 \times Z_2$, all of which have been widely used as flavor symmetries \cite{Dnpaper, D4papers, D3papers}.

Another way of obtaining discrete flavor symmetries from orbifolds is inspired by string theory and uses string selection rules \cite{strings}. We will not be using this approach and will only be employing regular field theory on an orbifold. However, as discussed in \cite{strings}, the two approaches do not contradict each other: If we have an orbifold possessing an inherent discrete symmetry, such as the ones we discuss in this paper, and then also impose the string selection rules, we will end up with an enlarged flavor symmetry.

This paper is organized as follows. In section \ref{possible} we discuss the possible 2-dimensional orbifolds and review how the discrete symmetries can be extracted from them. We also explain, why a 1-dimensional orbifold is not sufficient to obtain a non-abelian flavor symmetry. In section \ref{symmetrygroup} we discuss orbifold by orbifold which symmetry group arises from it. In section \ref{grouprep} we discuss the relation between flavor group representations and brane fields constrained to the fixed points in a certain twisted sector. Finally we conclude in section \ref{conclusions}.

\section{Orbifolding}
\label{possible}

We work in a  6-dimensional framework, where the two extra dimensions are compactified on an orbifold $T^2/Z_N$ \cite{ChoiKim}. The co-ordinates in the two extra dimensions are denoted by $(x_5,x_6)$.\\ \\
%%%%%%%%%%%%%%%The definition of torus%%%%%%%%%%%%%%%%%%%%%%
The $2$-dimensional torus $T^2$ is obtained by identifying the opposite sides of a parallelogram:

\begin{eqnarray}
(x_5,x_6) &\rightarrow& (x_5,x_6) + \vec{e}_1 \nonumber \\
(x_5,x_6) &\rightarrow& (x_5,x_6) + \vec{e}_2 \mbox{ } , \label{torus}
\end{eqnarray}
where $\vec{e}_1=(1,0), \mbox{   } \vec{e}_2=C(\cos{(\alpha)},\sin{(\alpha)})$ are the basis vectors of the torus. We can always choose $\vec{e}_1$ to point along the $x_5$ axis and to be normalized, leaving two free parameters defining $\vec{e}_2$, $C$ and $\alpha$, the length and the angle with respect to the $x_5$ axis. In this torus, the origin $(0,0)$ is identified with all points of the form

\begin{equation}
a \vec{e}_1 + b \vec{e}_2 \mbox{ } ,
\end{equation}

\noindent where $a,b$ are integers. 

%%%%%%%%%%%%%%%%The constraint of the moding out ZN group%%%

Aside from the torus basis, the orbifold is further defined by the abelian group $Z_N$ which is modded out of the torus. This means that we further identify points related by a rotation around the origin through integer multiples of an angle $\phi$, with $N \phi = 2 \pi$. The choice of $Z_N$ is strictly constrained, as we discuss in the following \cite{Lomont}. The group $Z_N$ is generated by one element, which corresponds to a rotation by the angle $\phi$. Its matrix representation in the Cartesian $x_5$-$x_6$ basis is thus 

\begin{equation}
\omega=\left( \begin{array}{cc} \cos{(\phi)} & -\sin{(\phi)} \\ \sin{(\phi)} & \cos{(\phi)}  \end{array} \right).
\end{equation}

Since the origin does not change under the rotation, all the points which are identified with the origin in the torus should be rotated to points which are also identified with the origin, i.e.

\begin{equation}
\omega (a \vec{e}_1 + b \vec{e}_2) = a' \vec{e}_1 + b' \vec{e}_2 \mbox{ } , 
\end{equation}
\noindent where $a, a', b$ and $b'$ are all integers.

Instead of using Cartesian coordinates, we can use the torus basis $\vec{e}_1,\vec{e}_2$. The matrix representation of the generating element in this basis reads 
\begin{equation}
\hat{\omega}=\left( \begin{array}{cc} n_1 & n_2 \\ n_3 & n_4  \end{array} \right),
\end{equation}
where $\hat{\omega}=U \omega U^{-1}$ and $U$ is the similarity transformation relating the Cartesian and Torus bases to each other. In this basis we have
\begin{equation}
\left( \begin{array}{cc} n_1 & n_2 \\ n_3 & n_4  \end{array} \right) \left( \begin{array}{c} a \\ b  \end{array} \right) = \left( \begin{array}{c} a' \\ b'  \end{array} \right).
\end{equation}
Due to the fact that $a,b,a',b'$ are integers, the $n_i$ must also be integers. And since the trace is a basis-independent quantity, we have
\begin{equation}
2 \cos{(\phi)} = Tr \omega = Tr \hat{\omega} = n_1 + n_4 \mbox{ } , 
\end{equation}
which implies that $2 \cos{(\phi)}$ is an integer and thus $\cos{(\phi)}=-1,-1/2,0,1/2,1$ corresponding to $\phi=\pi,2 \pi/3,\pi/2,\pi/3,2\pi$. This directly leads to a constraint for the $Z_N$, and we are only allowed to choose $N=1,2,3,4,6$. This then also leads to a constraint concerning our choice of torus basis vectors, since the rotational symmetry $Z_N$ needs to be consistent with the symmetry of the torus. When modding out $Z_2$, this is no constraint, as any basis is consistent with reflections. For $Z_3$ and $Z_6$ we can only take the relative angle between the basis vectors to be 60, 120 or 150 degrees. All three possibilities give the same orbifold. In this paper, we choose the $60^\circ$ lattice with basis vectors $(\vec{e}_1=(1,0),\vec{e}_2=(1/2,\sqrt{3}/2))$\footnote{The other two equivalent possibilities are the  $SU(3)$ lattice with $(\vec{e}_1=(1,0),\vec{e}_2=(-1/2,\sqrt{3}/2))$ and the $G_2$ lattice with $(\vec{e}_1=(1,0),\vec{e}_2=(-3/2,\sqrt{3}/2))$.}. Finally, when modding out $Z_4$ the only possibility is a $90^\circ$ lattice, with both basis vectors normalized to a length of 1.

We thus only have to discuss four different cases: $T^2/Z_2$, $T^2/Z_3$, $T^2/Z_4$ and $T^2/Z_6$. For the last three, the orbifold is uniquely defined, while for the first case we need to additionally discuss the effect of choosing a specific basis.

From these four orbifolds, we can then extract the residual Poincar\'e symmetry, which will in all cases be a  non-abelian discrete symmetry. This is done in the following way: After choosing the orbifold, we determine the fixed points. A fixed point is a point for which a rotation by an integer multiple of $\phi$ is equivalent to a lattice translation. These points are potential candidates for the localization of 3-branes \footnote{A 3-brane has three spatial dimensions.} and thus the Standard Model fermions can be taken to be brane fields, which are non-vanishing only at the fixed points. The fixed points are divided into several twisted sectors, where the $m$th twisted sector contains those fixed points for which a rotation by $m \phi$ corresponds to a lattice translation. A given fixed point can lie in several twisted sectors.

We assume all fixed points to be physically equivalent. This then means that the remnant translation and rotation symmetries are those which result only in a permutation of the fixed points, i.e. only map fixed points to other fixed points. These remnant symmetry operations are the elements of the residual Poincar\'e symmetry, and all that remains to be done is to find the underlying group structure.

One can then immediately see, why we do not need to consider the 1-dimensional orbifold $S^1/Z_N$: It has only two fixed points, and thus any symmetry group which permutes them will be a subgroup of the permutation group for two distinct objects, $S_2 \simeq Z_2$, which is abelian. Since we want to obtain a non-abelian discrete symmetry, we need to consider at least a 2-dimensional orbifold.

%%%%%%%%%%%%%%%%%%%%%%%%%%%%%%%%%%%%%%%%%%%%%%%%%%%%%%%%%%%%%%%%%%%%%%%%%%%%% 
\section{Symmetries from Orbifolding} \label{symmetrygroup}
%%%%%%%%%%%%%%%%%%%%%%%%%%%%%%%%%%%%%%%%%%%%%%%%%%%%%%%%%%%%%%%%%%%%%%%%%%%%%

In our discussion we parametrize the two extra dimensions by a complex number $z \equiv x_5+ ix_6$. Analogously to equation (\ref{torus}), the Torus $T^2$ is obtained by identifying the points in the complex plane related by
\begin{eqnarray}
z &\rightarrow& z+1 \mbox{ } , \\
z &\rightarrow& z+\gamma \mbox{ } , \label{torus2}
\end{eqnarray}

\noindent where the complex numbers $(1,\gamma)$ correspond to the basis vectors $(\vec{e}_1,\vec{e}_2)$.
%T2/Z2%
\subsection{$T^2/Z_2$}

\begin{figure}[t]
\epsfig{file=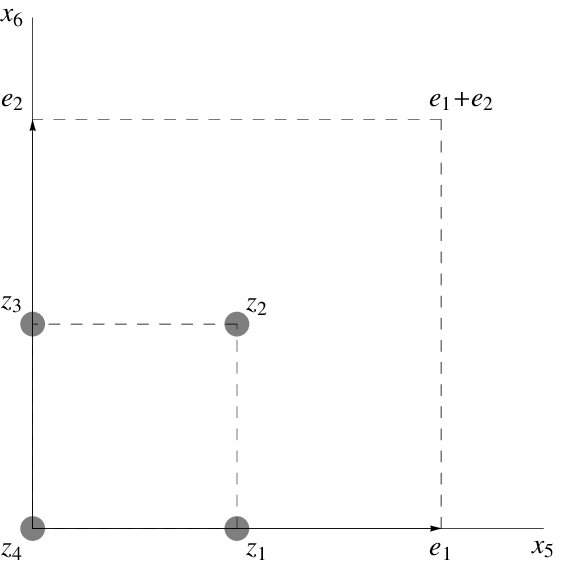,width=8cm,height=7cm}
\epsfig{file=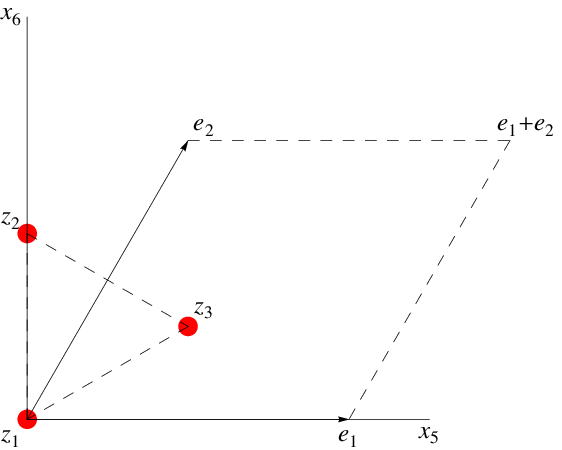,width=9cm,height=7cm}
\caption[]{\label{Z2andZ3}The orbifolds $T^2/Z_2$ (left) and $T^2/Z_3$ (right) with basis vectors $\vec{e_1}$, $\vec{e_2}$ and fixed points $z_i$. The square (left) and the triangle (right) formed by the fixed points, corresponding to the discrete symmetries $D_4$ and $D_3$ respectively, are shown with dashed lines.}
\end{figure}

If we mod out a $Z_2$ reflection symmetry, $\gamma$ can be arbitrary in general. However, in order to obtain a non-abelian symmetry, we have only two possibilities: The first one is $\gamma=e^{i \pi/3}$, which gives us an $S_4$ flavor symmetry, or an $A_4$ symmetry if only proper Lorentz transformations and translations (i.e. no discrete parities) are considered. The other possible basis is $\gamma=e^{i \pi/2}=i$. Since the case of $S_4$ and $A_4$ has already been discussed in \cite{GA}, we will only discuss the case $\gamma=e^{i \pi/2}=i$ here. This orbifold is shown in figure \ref{Z2andZ3}. The $Z_2$ parity is defined by
\begin{equation}
 z \rightarrow -z \mbox{ } .
\end{equation}
The fixed points are then given by $(z_1,z_2,z_3,z_4)=(1/2,(1+i)/2,i/2,0)$. The fixed points are permuted by the two translation operations
\begin{eqnarray}
S_1:z &\rightarrow& z + 1/2 \mbox{ } , \\
S_2:z &\rightarrow& z + i/2 \mbox{ } .  
\end{eqnarray}
Moreover, the fixed points are also permuted by the rotation
\begin{eqnarray}
T_R:z &\rightarrow& \omega z \mbox{ } ,
\end{eqnarray}
where $\omega=e^{i \pi/2}=i$. One can also write these operations explicitly in terms of the interchange of the fixed points,
\begin{eqnarray}
S_1[(14)(23)]:(z_1,z_2,z_3,z_4) &\rightarrow& (z_4,z_3,z_2,z_1) \mbox{ } , \\
S_2[(12)(34)]:(z_1,z_2,z_3,z_4) &\rightarrow& (z_2,z_1,z_4,z_3) \mbox{ } , \\
T_R[(13)(2)(4)]:(z_1,z_2,z_3,z_4) &\rightarrow& (z_3,z_2,z_1,z_4) \mbox{ } .
\end{eqnarray}
From these elements we can define two generators,
\begin{eqnarray}
A &=& [(13)(2)(4)][(14)(23)]=[(1432)] \mbox{ } , \\
B &=& [(12)(34)] \mbox{ } ,
\end{eqnarray}
satisfying the generator relations,
\begin{eqnarray}
A^4 &=& 1 \mbox{ } , \nonumber \\
B^2 &=& 1 \mbox{ } , \nonumber \\
ABA &=& B \mbox{ } . \label{D4algebra}
\end{eqnarray}
This describes the dihedral group $D_4$, the symmetry group of the square. The group theory of $D_4$, and of the dihedral groups in general, is discussed for example in \cite{Dnpaper}. Note that this group is not enlarged if we include parity transformations.

%%%%%%%%%%%%%%%%%%%%%%%%%%%%%%%%%%%%%%%%%%%%%%%%%%%%%%%%%%%%%%%%%%%%%%%%%%%%%%%%%%
%T2/Z3%

\subsection{$T^2/Z_3$}

When modding out $Z_3$ we consider, without loss of generality, only the torus with a $60^\circ$ lattice, as already mentioned in section \ref{possible}. This corresponds to the choice $\gamma=e^{i \pi/3}$. This orbifold is shown in figure \ref{Z2andZ3}. The operation of the generator of the $Z_3$  symmetry is given by
\begin{equation}
 z \rightarrow e^{i 2 \pi/3} z \mbox{ } .
\end{equation}
The corresponding fixed points are $(z_1,z_2,z_3)=(0,i/ \sqrt{3},1/2+i/ 2\sqrt{3})$. The translation operations permuting these fixed points are
\begin{eqnarray}
S_1:z &\rightarrow& z + (1/2+i/ 2\sqrt{3}) \mbox{ } , \\
S_2:z &\rightarrow& z + i/ \sqrt{3} \mbox{ } .
\end{eqnarray}
Moreover, the fixed points are also permuted by the rotation with respect to the origin
\begin{eqnarray}
T_R:z &\rightarrow& \omega z \mbox{ } , 
\end{eqnarray}
where $\omega=e^{i \pi/3}=i$. Again, one can also write the symmetry operations in terms of a permutation of the fixed points,
\begin{eqnarray}
S_1[(321)]:(z_1,z_2,z_3) &\rightarrow& (z_2,z_3,z_1) \mbox{ } , \\
S_2[(123)]:(z_1,z_2,z_3) &\rightarrow& (z_3,z_1,z_2) \mbox{ } , \\
T_R[(23)]:(z_1,z_2,z_3) &\rightarrow& (z_1,z_3,z_2) \mbox{ } .
\end{eqnarray}

A possible parity transformation would be equivalent to the rotation $T_R$ and thus does not need to be considered separately. We can formulate two generators
\begin{eqnarray}
A &=& [(321)] \mbox{ } , \\
B &=& [(321)][(23)]=[(13)] \mbox{ } ,
\end{eqnarray}
satisfying the generator relations,
\begin{eqnarray}
A^3 &=& 1 \mbox{ } , \nonumber \\
B^2 &=& 1 \mbox{ } , \nonumber \\
ABA &=& B \mbox{ } . \label{S3algebra}
\end{eqnarray}
This describes the dihedral group $D_3$, the symmetry group of the triangle, which is isomorphic to $S_3$ the permutation group of three distinct objects. As it is also a dihedral symmetry, its group theory is discussed in more detail in \cite{Dnpaper}.

%%%%%T2/Z4%%%%%%%%%%%%%%%

\subsection{$T^2/Z_4$}

\begin{figure}[t]
\epsfig{file=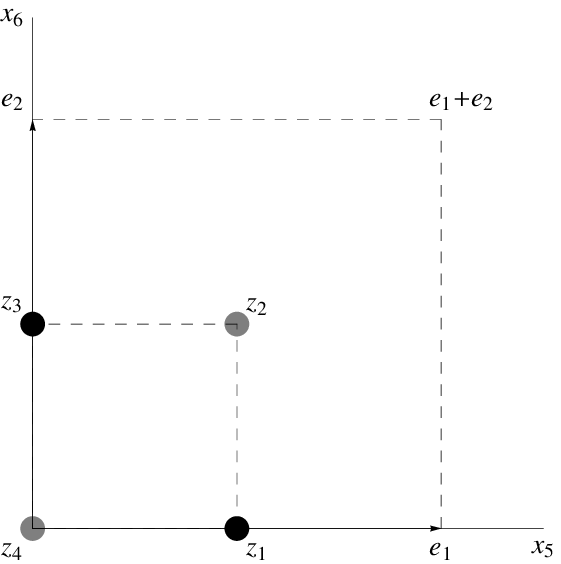,width=8cm,height=7cm}
\epsfig{file=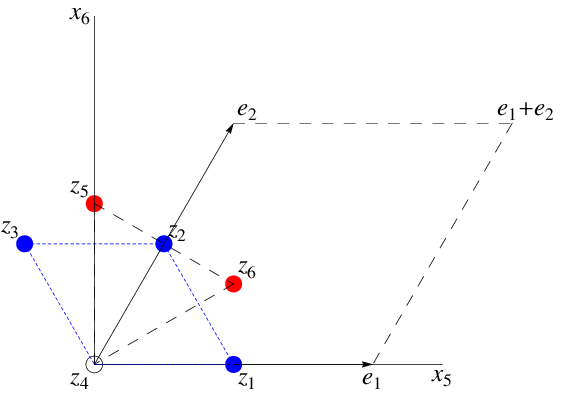,width=9cm,height=7cm}
\caption[]{\label{Z4andZ6} The orbifolds $T^2/Z_4$ (left) and $T^2/Z_6$ (right) with basis vectors $\vec{e_1}$, $\vec{e_2}$ and fixed points $z_i$. On the left, the fixed points which are both in the first and the second twisted sector are designated by gray points, those fixed points which are only in the second twisted sector are designated by black points. On the right, the fixed point which is in all twisted sectors is represented by a circle, those fixed points which are only in the second twisted sector are designated by red (lighter gray) points, while those fixed points which are only in the third twisted sector are given by blue (darker gray) points.}
\end{figure}

When modding out the abelian group $Z_4$, we have only one consistent choice of basis, $\gamma=e^{i \pi/2}=i$. The torus is the same one we used for $T^2/Z_2$ to obtain the $D_4$ symmetry, as one can also see in figure \ref{Z4andZ6}. In fact, the fixed points will also be the same and we thus obtain the same flavor symmetry. This is due to the fact that we obtain all fixed points of the orbifold $T^2/Z_4$ in the second twisted sector, where we only consider the squared generator of $Z_4$. This corresponds to a $Z_2$ subgroup of $Z_4$ and is thus fully equivalent to our discussion for $T^2/Z_2$ with a $90^{\circ}$ lattice. The first twisted sector only contains the fixed points $z_2$ and $z_4$; as both of them also appear in the second twisted sector no new fixed points and thus no new residual translational or rotational symmetry operations arise due to the larger abelian group, $Z_4$. The unique symmetry we thus obtain is $D_4$.

%%%%%%%%%T2/Z6%%%%%%%%%%%

\subsection{$T^2/Z_6$}

As for $T^2/Z_3$ we use the $60^\circ$ lattice, i.e. $\gamma=e^{i \pi/3}$. The orbifold is shown in figure \ref{Z4andZ6}. The operation of the $Z_6$ symmetry for the first twisted sector is defined by
\begin{equation}
 z \rightarrow e^{i 2\pi/6} z \mbox{ } .
\end{equation}
For the first twisted sector, we have only one fixed point which is $z_4=0$. For the second twisted sector, the operation of the $Z_6$ symmetry reads
\begin{equation}
 z \rightarrow e^{i 2\pi/3} z \mbox{ } .
\end{equation}
The fixed points of the second twisted sector are $(z_4,z_5,z_6)=(0,i/ \sqrt{3},1/2(1+i/ \sqrt{3})$ which are the same as in the case of $T^2/Z_3$.\\
For the third twisted sector, finally, the operation of $Z_6$ symmetry is written as
\begin{equation}
 z \rightarrow -z \mbox{ } .
\end{equation}
The fixed points in this sector thus are $(z_1,z_2,z_3,z_4)=(1/2,1/4+ i \sqrt{3}/4,-1/4+ i \sqrt{3}/4,0)$. Combining all fixed points $(z_1,z_2,z_3,z_4,z_5,z_6)$, we find that the fixed points are only permuted by residual rotation operations, i.e. translation symmetry is fully broken. These rotations are
\begin{eqnarray}
T_{R1}:z &\rightarrow& e^{i \pi/3} z \mbox{ } , \\
T_{R2}:z &\rightarrow& e^{i 2 \pi/3} z \mbox{ } .  
\end{eqnarray}
Moreover, if we assume the full Poincar\'e symmetry, we also have two parity operations acting on the fixed points
\begin{eqnarray}
P_1: z &\rightarrow& z^{*}, \\
P_2: z &\rightarrow& -z^{*},   
\end{eqnarray}
where $z^{*}$ denotes the complex conjugation of $z$. 

We can write all of these symmetry operations in terms of a permutation of the fixed points as
\begin{eqnarray}
T_{R1}[(123)(56)]:(z_1,z_2,z_3,z_4,z_5,z_6) &\rightarrow& (z_3,z_1,z_2,z_4,z_6,z_5) \mbox{ } , \\
T_{R2}[(132)]:(z_1,z_2,z_3,z_4,z_5,z_6) &\rightarrow& (z_2,z_3,z_1,z_4,z_5,z_6) \mbox{ } , \\
P_1[(23)(56)]:(z_1,z_2,z_3,z_4,z_5,z_6) &\rightarrow& (z_1,z_3,z_2,z_4,z_6,z_5) \mbox{ } , \\
P_2[(23)]:(z_1,z_2,z_3,z_4,z_5,z_6) &\rightarrow& (z_1,z_3,z_2,z_4,z_5,z_6) \mbox{ } .
\end{eqnarray}
From these operators, we can form the generators
\begin{eqnarray}
A &=& [(123)(56)] \mbox{ } , \\
B &=& [(23)] \mbox{ } ,
\end{eqnarray}
satisfying the generator relations,
\begin{eqnarray}
A^6 &=& 1 \mbox{ } , \nonumber \\
B^2 &=& 1 \mbox{ } , \nonumber \\
ABA &=& B \mbox{ } . \label{D6algebra}
\end{eqnarray}
This defines the group $D_6 \cong D_3 \times Z_2 \cong S_3 \times Z_2$. If we do not include the parity operations, we effectively lose the generator $B$. The flavor symmetry then has only one generator and is the abelian group $Z_6$.

%%%%%%%%%%%%%%%%%%%%%%%%%%%%%%%%%%%%%%%%%%%%%%%

\section{Group Representations} \label{grouprep}

To construct a full model, one now needs to assign the fermion generations to representations of these flavor groups. The orbifold fixed points are interpreted as 3-branes, on which the fermion fields are localized. The flavor symmetry operations which permute the fixed points then act non-trivially on the fermion fields. Irreducible representations correspond to relations among the field values at different fixed points; these relations are invariant under symmetry operations. In general this means that one or more fermion generations transforming under an irreducible representation of the flavor group will be ``smeared out'' over all available fixed points. All representations can be reproduced in this way, the origin of the flavor group from orbifolding thus does not offer any restrictions on the choice of representations. Also, all representations will correspond in general to the field(s) being non-vanishing at all fixed points. Thus, although the flavor symmetry as a whole has a straightforward interpretation in the geometry of the orbifold, the different representations do not.

This is at least a bit different for the last orbifold we have discussed, $T^2/Z_6$. The resulting flavor symmetry was $D_6$, which is isomorphic to $D_3 \times Z_2$. We observe that all symmetry operations leave the origin, the fixed point $z_4$, invariant. Thus a field which is localized at the origin will transform trivially under the flavor symmetry. In addition the subgroup $D_3$ generated by $A^2$ and $B$ leaves the fixed points $z_5$ and $z_6$, i.e. the fixed points of the second twisted sector, invariant. Fields localized only on these two fixed points thus transform non-trivially only under the $Z_2$ factor of the flavor group. Similarly, the fixed points of the third twisted sector, $z_1, z_2$ and $z_3$ are not permuted by the group element $A^3$, which generates $Z_2$. Fields localized in this sector will thus only transform non-trivially under the $D_3$ factor of the flavor group. Fields transforming non-trivially both under $D_3$ and $Z_2$ will necessarily be non-vanishing in both the second and the third twisted sector. For more details on the representation theory of $D_6$ and the transformation properties of representations under subgroups, see \cite{Dnpaper}.

The orbifold $T^2/Z_6$ thus offers the aesthetical appeal that different representations correspond to different localizations in the orbifold and therefore have a more intuitive interpretation in terms of the orbifold geometry. However also here all representations can be reproduced, and the orbifold origin of the flavor symmetry does not offer further input as to which representations to use for model building.

%%%%%%%%%%%%%%%%%%%%%%%%%%%%%%%%%%%%%%%%%%%%%%%
\section{Conclusion}
\label{conclusions}

We have discussed all possible non-abelian discrete symmetries arising from 2-dimensional orbifolds. In this context the flavor symmetries arise as a remnant symmetry of the full 6-dimensional space-time symmetry. This remnant symmetry can then be interpreted as the permutation symmetry of the orbifold fixed points. These fixed points in turn are taken to be 3-branes, on which the three generations of Standard Model fermions reside. The flavor symmetry then has a straightforward interpretation in terms of the geometry of the orbifold. As in crystallography, the number of possible lattice structures and symmetry groups is strictly limited for orbifolds. The resulting flavor symmetries are all crystallographic point groups, as was to be expected. The possible flavor groups we obtain are $S_4,A_4,S_3,D_4,$ and $D_6 \simeq D_3 \times Z_2$, where the first two were already discussed in \cite{GA}. All of these groups have been widely used as phenomenologically successful flavor symmetries.

The question is of course what implications these findings have for flavored model building. We found that, in all cases, the representation content of such models is not limited by assuming an orbifold origin for the flavor symmetry. In the case of $D_6$ at least the representations can be interpreted in terms of localization of the fields in specific sectors, but without a full theory of why certain fields are localized where, this does not offer direct model building input either. One can thus take two major hints from this general approach: First of all, it favors the well-known, small crystallographic groups as flavor symmetries, due to the crystalline structure of the two extra dimensions. This is however due to the fact that we have only considered two extra dimensions. Larger, more complicated, flavor symmetries may arise when considering more extra dimensions. This leads us to the second point: Further input for model building requires an extended analysis of the extra-dimensional setup. This has been done in the context of string theory \cite{strings}, where interestingly enough the flavor group $D_4$ also appears naturally. However, the results of this paper can also be combined with regular extra-dimensional field theory. For example, the flavor groups arising from the orbifolds need to be further broken, a process which may actually be intimately connected with the extra dimensions themselves \cite{FlavorVEV}.

\subsubsection*{Acknowledgements}

We would like to thank Claudia Hagedorn and Stefan Groot Nibbelink for useful discussions. A.B. acknowledges support from the Studienstiftung des deutschen Volkes.

%Bibliography%


\begin{thebibliography}{99}

%\cite{Altarelli:2006kg}
\bibitem{GA}
  G.~Altarelli, F.~Feruglio and Y.~Lin,
  %``Tri-bimaximal neutrino mixing from orbifolding,''
  Nucl.\ Phys.\  B {\bf 775}, 31 (2007)
  [arXiv:hep-ph/0610165].
  %%CITATION = NUPHA,B775,31;%%

\bibitem{nogo}
  A.~Adulpravitchai, A.~Blum and M.~Lindner,
   in preparation.
  


\bibitem{A4papers}

E.~Ma and G.~Rajasekaran,
%  {\it Softly broken A(4) symmetry for nearly degenerate neutrino masses},
  Phys.\ Rev.\ D {\bf 64}, 113012 (2001)
  [arXiv:hep-ph/0106291];
K.~S.~Babu, E.~Ma and J.~W.~F.~Valle,
%  {\it Underlying A(4) symmetry for the neutrino mass matrix and the quark  mixing matrix},
  Phys.\ Lett.\ B {\bf 552}, 207 (2003)
  [arXiv:hep-ph/0206292];
M.~Hirsch, J.~C.~Romao, S.~Skadhauge, J.~W.~F.~Valle and A.~Villanova del Moral,
%  {\it Degenerate neutrinos from a supersymmetric A(4) model},
  arXiv:hep-ph/0312244;
%  {\it Phenomenological tests of supersymmetric A(4) family symmetry model of neutrino mass},
  Phys.\ Rev.\  D {\bf 69}, 093006 (2004)
  [arXiv:hep-ph/0312265];
E.~Ma,
%  {\it Quark mass matrices in the A(4) model},
  Mod.\ Phys.\ Lett.\ A {\bf 17}, 627 (2002)
  [arXiv:hep-ph/0203238];
  %%CITATION = HEP-PH 0203238;%%
%  {\it A(4) origin of the neutrino mass matrix},
  Phys.\ Rev.\ D {\bf 70}, 031901 (2004)
  [arXiv:hep-ph/0404199];
%  {\it Non-Abelian discrete symmetries and neutrino masses: Two examples},
  New J.\ Phys.\  {\bf 6}, 104 (2004) 
  [arXiv:hep-ph/0405152];
%  {\it Non-Abelian discrete family symmetries of leptons and quarks},
  arXiv:hep-ph/0409075;
  %%CITATION = HEP-PH/0409075;%%
%  {\it Aspects of the tetrahedral neutrino mass matrix},
  Phys.\ Rev.\  D {\bf 72}, 037301 (2005)
  [arXiv:hep-ph/0505209];
%  {\it Tetrahedral family symmetry and the neutrino mixing matrix},
  Mod.\ Phys.\ Lett.\ A {\bf 20}, 2601 (2005)
  [arXiv:hep-ph/0508099];
%  {\it Tribimaximal neutrino mixing from a supersymmetric model with A4 family symmetry},
  Phys.\ Rev.\  D {\bf 73}, 057304 (2006) 
  [arXiv:hep-ph/0511133];
%  {\it Suitability of A(4) as a family symmetry in grand unification},
  Mod.\ Phys.\ Lett.\  A {\bf 21},  2931(2006)
  [arXiv:hep-ph/0607190];
%  {\it Supersymmetric A(4) x Z(3) and A(4) realizations of neutrino  tribimaximal mixing without and with corrections},
  Mod.\ Phys.\ Lett.\  A {\bf 22}, 101 (2007)
  [arXiv:hep-ph/0610342];
S.~L.~Chen, M.~Frigerio and E.~Ma,
%  {\it Hybrid seesaw neutrino masses with A(4) family symmetry},
  Nucl.\ Phys.\  B {\bf 724}, 423 (2005)
  [arXiv:hep-ph/0504181];
  %%CITATION = NUPHA,B724,423;%%
K.~S.~Babu and X.~G.~He,
%  {\it Model of geometric neutrino mixing},
  arXiv:hep-ph/0507217;
A.~Zee,
% {\it Obtaining the neutrino mixing matrix with the tetrahedral group},
  Phys.\ Lett.\ B {\bf 630}, 58 (2005)
  [arXiv:hep-ph/0508278];
G.~Altarelli and F.~Feruglio,
  %{\it Tri-bimaximal neutrino mixing from discrete symmetry in extra dimensions},
  Nucl.\ Phys.\ B {\bf 720}, 64 (2005)
  [arXiv:hep-ph/0504165];
  %{\it Tri-bimaximal neutrino mixing, A(4) and the modular symmetry},
  Nucl.\ Phys.\ B {\bf 741}, 215 (2006)
  [arXiv:hep-ph/0512103];
 G.~Altarelli, F.~Feruglio and Y.~Lin,
  %{\it Tri-bimaximal neutrino mixing from orbifolding},
  Nucl.\ Phys.\  B {\bf 775}, 31 (2007)
  [arXiv:hep-ph/0610165];
  %%CITATION = NUPHA,B775,31;%%
 G.~Altarelli, F.~Feruglio and C.~Hagedorn,
  %{\it A SUSY SU(5) Grand Unified Model of Tri-Bimaximal Mixing from A(4)},
  JHEP {\bf 0803}, 052 (2008)
  [arXiv:0802.0090 [hep-ph]];
 Y.~Lin,
  % {\it A predictive A4 model, Charged Lepton Hierarchy and Tri-bimaximal Sum Rule},
  arXiv:0804.2867 [hep-ph].
X.~G.~He, Y.~Y.~Keum and R.~R.~Volkas,
%  {\it A(4) flavour symmetry breaking scheme for understanding quark and  neutrino mixing angles},
  JHEP {\bf 0604}, 039 (2006)
  [arXiv:hep-ph/0601001];
B.~Adhikary, B.~Brahmachari, A.~Ghosal, E.~Ma and M.~K.~Parida,
%  {\it A(4) symmetry and prediction of U(e3) in a modified Altarelli-Feruglio model},
  Phys.\ Lett.\ B {\bf 638}, 345 (2006)
  [arXiv:hep-ph/0603059];
L.~Lavoura and H.~Kuhbock,
%  {\it Predictions of an A(4) model with a five-parameter neutrino mass matrix},
  Mod.\ Phys.\ Lett.\  A {\bf 22}, 181 (2007)
  [arXiv:hep-ph/0610050];
S.~F.~King and M.~Malinsky,
%  {\it A(4) family symmetry and quark-lepton unification},
  Phys.\ Lett.\  B {\bf 645}, 351 (2007)
  [arXiv:hep-ph/0610250];
S.~Morisi, M.~Picariello and E.~Torrente-Lujan,
%  {\it A model for fermion masses and lepton mixing in SO(10) x A(4)},
  Phys.\ Rev.\  D {\bf 75}, 075015 (2007)
  [arXiv:hep-ph/0702034];
F.~Yin,
% {\it Neutrino mixing matrix in the 3-3-1 model with heavy leptons and A(4) symmetry},
  Phys.\ Rev.\  D {\bf 75}, 073010 (2007)
  [arXiv:0704.3827 [hep-ph]];
F.~Bazzocchi, S.~Kaneko and S.~Morisi,
%  {\it A SUSY A(4) model for fermion masses and mixings},
  JHEP {\bf 0803}, 063 (2008)
  [arXiv:0707.3032 [hep-ph]];
F.~Bazzocchi, S.~Morisi and M.~Picariello,
%  {\it Embedding A(4) into left-right flavor symmetry: Tribimaximal neutrino mixing and fermion hierarchy},
  Phys.\ Lett.\  B {\bf 659}, 628 (2008)
  [arXiv:0710.2928 [hep-ph]];
M.~Honda and M.~Tanimoto,
%  {\it Deviation from tri-bimaximal neutrino mixing in A(4) flavor symmetry},
  Prog.\ Theor.\ Phys.\  {\bf 119}, 583 (2008)
  [arXiv:0801.0181 [hep-ph]];
B.~Brahmachari, S.~Choubey and M.~Mitra,
%  {\it The A(4) flavor symmetry and neutrino phenomenology},
  Phys.\ Rev.\  D {\bf 77}, 073008 (2008)
  [Erratum-ibid.\  D {\bf 77}, 119901 (2008)]
  [arXiv:0801.3554 [hep-ph]];
F.~Bazzocchi, S.~Morisi, M.~Picariello and E.~Torrente-Lujan,
  %``Embedding A4 into SU(3)xU(1) flavor symmetry: Large neutrino mixing and
  %fermion mass hierarchy in SO(10) GUT,''
  J.\ Phys.\ G {\bf 36}, 015002 (2009)
  [arXiv:0802.1693 [hep-ph]];
P.~H.~Frampton and S.~Matsuzaki,
%  {\it Renormalizable $A_4$ Model for Lepton Sector},
  arXiv:0806.4592 [hep-ph].
M.~C.~Chen and S.~F.~King,
  %``$A_4$ See-Saw Models and Form Dominance,''
  arXiv:0903.0125 [hep-ph].
  %%CITATION = ARXIV:0903.0125;%%

\bibitem{S4papers}
 C.~Hagedorn, M.~Lindner and R.~N.~Mohapatra,
  %``S(4) flavor symmetry and fermion masses: Towards a grand unified theory  of
  %flavor,''
  JHEP {\bf 0606}, 042 (2006)
  [arXiv:hep-ph/0602244].
  %%CITATION = JHEPA,0606,042;%%
C.~S.~Lam,
  %``Determining Horizontal Symmetry from Neutrino Mixing,''
  Phys.\ Rev.\ Lett.\  {\bf 101}, 121602 (2008)
  [arXiv:0804.2622 [hep-ph]];
  %%CITATION = PRLTA,101,121602;%%
  %``The Unique Horizontal Symmetry of Leptons,''
  Phys.\ Rev.\  D {\bf 78}, 073015 (2008)
  [arXiv:0809.1185 [hep-ph]];
  %%CITATION = PHRVA,D78,073015;%%
  F.~Bazzocchi and S.~Morisi,
  %``S4 as a natural flavor symmetry for lepton mixing,''
  arXiv:0811.0345 [hep-ph].
  %%CITATION = ARXIV:0811.0345;%%
G.~Altarelli, F.~Feruglio and L.~Merlo,
  %``Revisiting Bimaximal Neutrino Mixing in a Model with S4 Discrete
  %Symmetry,''
  arXiv:0903.1940 [hep-ph].
  %%CITATION = ARXIV:0903.1940;%%
H.~Ishimori, Y.~Shimizu and M.~Tanimoto,
  %``S4 Flavor Symmetry of Quarks and Leptons in SU(5) GUT,''
  arXiv:0812.5031 [hep-ph].
  %%CITATION = ARXIV:0812.5031;%%

\bibitem{Dnpaper}
  A.~Blum, C.~Hagedorn and M.~Lindner,
  %``Fermion masses and mixings from dihedral flavor symmetries with preserved subgroups,''
  Phys.\ Rev.\  D {\bf 77}, 076004 (2008)
  [arXiv:0709.3450v1 [hep-ph]].

\bibitem{D4papers}
W.~Grimus and L.~Lavoura,
  %``A discrete symmetry group for maximal atmospheric neutrino mixing,''
  Phys.\ Lett.\  B {\bf 572}, 189 (2003)
  [arXiv:hep-ph/0305046].
  %%CITATION = PHLTA,B572,189;%%
W.~Grimus, A.~S.~Joshipura, S.~Kaneko, L.~Lavoura and M.~Tanimoto,
  %``Lepton mixing angle theta(13) = 0 with a horizontal symmetry D(4),''
  JHEP {\bf 0407}, 078 (2004)
  [arXiv:hep-ph/0407112].
  %%CITATION = JHEPA,0407,078;%%
H.~Ishimori, T.~Kobayashi, H.~Ohki, Y.~Omura, R.~Takahashi and M.~Tanimoto,
  %``D4 Flavor Symmetry for Neutrino Masses and Mixing,''
  Phys.\ Lett.\  B {\bf 662}, 178 (2008)
  [arXiv:0802.2310 [hep-ph]].
  %%CITATION = PHLTA,B662,178;%%
H.~Ishimori, T.~Kobayashi, H.~Ohki, Y.~Omura, R.~Takahashi and M.~Tanimoto,
  %``Soft supersymmetry breaking terms from D4 x Z2 lepton flavor symmetry,''
  Phys.\ Rev.\  D {\bf 77}, 115005 (2008)
  [arXiv:0803.0796 [hep-ph]].
  %%CITATION = PHRVA,D77,115005;%%
A.~Adulpravitchai, A.~Blum and C.~Hagedorn,
  %``A Supersymmetric D4 Model for mu-tau Symmetry,''
  JHEP {\bf 0903}, 046 (2009)
  [arXiv:0812.3799 [hep-ph]].
  %%CITATION = JHEPA,0903,046;%%


\bibitem{D3papers}
E.~Ma,
  %``Permutation symmetry for neutrino and charged-lepton mass matrices,''
  Phys.\ Rev.\  D {\bf 61}, 033012 (2000)
  [arXiv:hep-ph/9909249].
  %%CITATION = PHRVA,D61,033012;%%
J.~Kubo, A.~Mondragon, M.~Mondragon and E.~Rodriguez-Jauregui,
  %``The flavor symmetry,''
  Prog.\ Theor.\ Phys.\  {\bf 109}, 795 (2003)
  [Erratum-ibid.\  {\bf 114}, 287 (2005)]
  [arXiv:hep-ph/0302196].
  %%CITATION = PTPKA,109,795;%%
S.~L.~Chen, M.~Frigerio and E.~Ma,
  %``Large neutrino mixing and normal mass hierarchy: A discrete
  %understanding,''
  Phys.\ Rev.\  D {\bf 70}, 073008 (2004)
  [Erratum-ibid.\  D {\bf 70}, 079905 (2004)]
  [arXiv:hep-ph/0404084].
  %%CITATION = PHRVA,D70,073008;%%
 W.~Grimus and L.~Lavoura,
  %``S(3) x Z(2) model for neutrino mass matrices,''
  JHEP {\bf 0508}, 013 (2005)
  [arXiv:hep-ph/0504153].
  %%CITATION = JHEPA,0508,013;%%
N.~Haba and K.~Yoshioka,
  %``Discrete flavor symmetry, dynamical mass textures, and grand
  %unification,''
  Nucl.\ Phys.\  B {\bf 739}, 254 (2006)
  [arXiv:hep-ph/0511108].
  %%CITATION = NUPHA,B739,254;%%
  T.~Teshima,
  %``Flavor mass and mixing and S(3) symmetry: An S(3) invariant model
  %reasonable to all,''
  Phys.\ Rev.\  D {\bf 73}, 045019 (2006)
  [arXiv:hep-ph/0509094].
  %%CITATION = PHRVA,D73,045019;%%
F.~Caravaglios and S.~Morisi,
  %``Neutrino masses and mixings with an S(3) family permutation symmetry,''
  arXiv:hep-ph/0503234.
  %%CITATION = HEP-PH/0503234;%%
L.~Lavoura and E.~Ma,
  %``Two predictive supersymmetric S(3) x Z(2) models for the quark mass
  %matrices,''
  Mod.\ Phys.\ Lett.\  A {\bf 20}, 1217 (2005)
  [arXiv:hep-ph/0502181].
  %%CITATION = MPLAE,A20,1217;%%
S.~Kaneko, H.~Sawanaka, T.~Shingai, M.~Tanimoto and K.~Yoshioka,
  %``Flavor Symmetry and Vacuum Aligned Mass Textures,''
  Prog.\ Theor.\ Phys.\  {\bf 117}, 161 (2007)
  [arXiv:hep-ph/0609220].
  %%CITATION = PTPKA,117,161;%%
R.~N.~Mohapatra, S.~Nasri and H.~B.~Yu,
  %``S(3) symmetry and tri-bimaximal mixing,''
  Phys.\ Lett.\  B {\bf 639}, 318 (2006)
  [arXiv:hep-ph/0605020].
  %%CITATION = PHLTA,B639,318;%%
Y.~Kajiyama, J.~Kubo and H.~Okada,
  %``D6 Family Symmetry and Cold Dark Matter at LHC,''
  Phys.\ Rev.\  D {\bf 75}, 033001 (2007)
  [arXiv:hep-ph/0610072].
  %%CITATION = PHRVA,D75,033001;%%
C.~D.~Carone and R.~F.~Lebed,
  %``A hexagonal theory of flavor,''
  Phys.\ Rev.\  D {\bf 60}, 096002 (1999)
  [arXiv:hep-ph/9905275].
  %%CITATION = PHRVA,D60,096002;%%
N.~Haba, A.~Watanabe and K.~Yoshioka,
  %``Twisted flavors and tri/bi-maximal neutrino mixing,''
  Phys.\ Rev.\ Lett.\  {\bf 97}, 041601 (2006)
  [arXiv:hep-ph/0603116].
  %%CITATION = PRLTA,97,041601;%%
F.~Feruglio and Y.~Lin,
  %``Fermion Mass Hierarchies and Flavour Mixing from a Minimal Discrete
  %Symmetry,''
  Nucl.\ Phys.\  B {\bf 800}, 77 (2008)
  [arXiv:0712.1528 [hep-ph]].
  %%CITATION = NUPHA,B800,77;%%
  
  %\cite{Kobayashi:2006wq}
\bibitem{strings}
  T.~Kobayashi, H.~P.~Nilles, F.~Ploger, S.~Raby and M.~Ratz,
  %``Stringy origin of non-Abelian discrete flavor symmetries,''
  Nucl.\ Phys.\  B {\bf 768}, 135 (2007)
  [arXiv:hep-ph/0611020].
  %%CITATION = NUPHA,B768,135;%%


\bibitem{ChoiKim}
  Kang-Sin Choi and Jihn E. Kim,
  %``Quarks and Leptons From Orbifolded Superstring,''
  Lect. Notes Phys. {\bf 696}, Springer, Berlin Heidelberg 2006.

\bibitem{Lomont}
  J.~S.~Lomont,
  %``Applications of Finite Groups,''
  Dover publication, INC, New York, 1993.

\bibitem{FlavorVEV}
  T.~Kobayashi, Y.~Omura, K.~Yoshioka,
  %``Flavor Symmetry Breaking and Vacuum Alignment on Orbifolds,''
  Phys.\ Rev.\ D {\bf 78}, 115006, (2008)
  [arXiv:0809.3064 [hep-ph]]
 
\end{thebibliography}
\end{document}